# Anomalous current-induced spin torques in ferrimagnets near compensation

Rahul Mishra[1], Jiawei Yu[1], Xuepeng Qiu[2], M. Motapothula[3], T. Venkatesan[1,3,4,5,6], and Hyunsoo Yang[1,3*]

[1]Department of Electrical and Computer Engineering, National University of Singapore, 117576, Singapore

[2]Shanghai Key Laboratory of Special Artificial Microstructure Materials & School of Physics Science and Engineering, Tongji University, Shanghai 200092, China

[3]NUSNNI-NanoCore, National University of Singapore, 117411, Singapore

[4]Department of Physics, National University of Singapore, Singapore 117542, Singapore

[5]Department of Materials Science and Engineering, National University of Singapore, Singapore 117542, Singapore

[6]Integrated Science and Engineering Department, National University of Singapore, Singapore 117542, Singapore

While current-induced spin-orbit torques (SOTs) have been extensively studied in ferromagnets and antiferromagnets, ferrimagnets have been less studied. Here we report the presence of enhanced spin-orbit torques resulting from negative exchange interaction in ferrimagnets. The effective field and switching efficiency increase substantially as CoGd approaches its compensation point, giving rise to 9 times larger spin-orbit torques compared to that of non-compensated one. The macrospin modelling results also support efficient spin-orbit torques in a ferrimagnet. Our results suggest that ferrimagnets near compensation can be a new route for spin-orbit torque applications due to their high thermal stability and easy current-induced switching assisted by negative exchange interaction.

*eleyang@nus.edu.sg

SOTs have emerged as proficient means of manipulating the magnetization [1–9], in which the spin current generated from a heavy metal transfers its angular momentum to the adjacent magnet. A wide range of SOT experiments have demonstrated that the spin current can be modulated by materials [10–12], structural [13–15] and interface [16–18] engineering. On the other hand, a magnet itself can also play an important role in modulating the spin torques as suggested by recent experiments and theory [19–23]. A majority of these reports use magnetic layers with an antiferromagnetic ordering.

A key characteristic of the materials with antiferromagnetic ordering is the strong negative exchange interaction. Due to the negative exchange, antiferromagnets and ferrimagnets show features distinct to their ferromagnetic counterparts. For example, the time scale of magnetization dynamics and reversal for antiferromagnets and ferrimagnets (ps) is much lower compared to that of ferromagnets (ns) [19,24,25]. Similarly, a faster demagnetization is observed in the case of antiparallel coupling as compared to parallel coupling in Co/Pt multilayers separated by a Ru spacer [26]. Negative exchange coupling is also found to assist in efficient domain wall motion in synthetic antiferromagnets [20]. In light of the above observations, an active role of antiferromagnetic exchange is expected in modulating SOTs in magnetic switching devices. Ferrimagnets provide an ideal platform to explore this possibility due to their tunable exchange interaction [27]. Recent SOT studies on collinear ferrimagnets were limited to one or two fixed compositions [28,29]. However, in order to evaluate the effect of exchange interaction on SOTs, a comparative study across different ferrimagnetic compositions is required.

In this Letter, we demonstrate that the negative exchange interaction torque can enhance the SOTs in a thermally stable, thick (6 nm) collinear ferrimagnet, CoGd. Unlike a ferromagnetic system, the SOT effective fields in Pt/CoGd are found not to scale inversely with $M_s$. As the ferrimagnet approaches compensation, the longitudinal SOT effective field ($H_L$) and switching efficiency ($\chi$) increase ~ 9 and 6 times, respectively, while the decrease in $M_S$ is only two fold. The anomalous increase of the SOT efficiency ($H_L$ and $\chi$) near the compensation is attributed to the presence of an additional torque in ferrimagnets which increases as the ferrimagnet approaches compensation. This additional torque which we refer to as exchange interaction torque is due to the negative exchange interaction between the ferrimagnetic sub-lattices.

CoGd is a rare earth-transition metal (RE-TM) ferrimagnet in which the Co and Gd are coupled antiferromagnetically [27,30]. Therefore, the magnetization of CoGd can be tuned to be dominated either by Co or Gd depending on their relative composition. At a magnetic compensation point, the total Co and Gd moments are equal, so the net magnetic moment of the CoGd system is zero [27]. CoGd also has a bulk perpendicular magnetic anisotropy (PMA) for various compositions depending on the deposition parameters [27,30,31], and consequently a thick magnetic layer can be grown. In spite of almost zero saturation magnetization ($M_s$) near compensation, CoGd has a finite spin polarization which results in a non-zero tunnel magnetoresistance (TMR) in magnetic tunnel junctions [32], which facilitates the reading operation of ferrimagnet based memory devices.

The film stacks of Si substrate/Pt (10 nm)/$Co_{1-x}Gd_x$ (6 nm)/$TaO_x$ (1 nm) were deposited on thermally oxidized Si substrates using magnetron sputtering with a base pressure less than $5 \times 10^{-9}$ Torr. The CoGd layer was deposited by co-sputtering from Co and Gd targets. The sputtering power of Co target was fixed at 120 W while varying the sputtering power of Gd target from 60 to 120 W. The Co and Gd composition in the films was determined to be in the range from $Co_{83}Gd_{17}$ to $Co_{64}Gd_{36}$ by Rutherford backscattering spectrometry. All the films showed PMA, confirmed by polar magneto-optic Kerr effect (MOKE) measurements. Subsequently the films were patterned into Hall bar devices with a width of 10 µm, using photolithography and ion-milling. A thick Pt channel (10 nm) was used to ensure that the resistance of the patterned devices does not vary widely and the current distribution profile remains similar for different samples.

Figure 1(a) shows the schematic of the transport measurement geometry. Anomalous Hall measurements were performed by sweeping the magnetic field out of plane and measuring the Hall voltage. The samples with $x$ (Gd % in CoGd) ranging from 20 to 24.2 were identified to be Co dominated using the anomalous Hall resistance, $R_{AHE}$ (Fig. 1(b)) and MOKE (inset of Fig. 1(c)) measurements, while samples with $x$ ranging from 25.5 to 34 were Gd dominated [33,34]. A negative $R_{AHE}$ polarity for Gd rich samples can be attributed to the fact that the $R_{AHE}$ is dominated by Co in the CoGd ferrimagnetic system [35,36]. Figure 1(c) shows the value of coercivity for different $x$, obtained from the MOKE and $R_{AHE}$ measurements. As it is known that the coercivity diverges and reaches its peak near compensation for RE-TM alloys [30,37], there is a peak at $x \sim 25$. Near compensation, the $M_s$ is minimum, measured by a vibrating sample magnetometer (VSM), as shown in Figure 1(d). Throughout the above material characterization, we identify $x \sim 25$ as a crossover composition between a Co to Gd rich state.

We have then performed harmonic (Fig. 2(a,b)) and SOT switching measurements (Fig. 2(c,d)) on patterned Hall bar devices to determine current-induced effective fields and SOT switching efficiency ($\chi$), respectively. The data from two representative samples for Co and Gd rich regime are shown in Fig. 2. The second harmonic ac measurements were performed to evaluate the longitudinal ($H_L$) and transverse ($H_T$) effective fields [5,6,15,38–40]. Low frequency (13.7 Hz) ac currents with an amplitude of 10 mA ($6.25 \times 10^{10}$ A/m$^2$) was passed through the Hall bar. The 1$^{st}$ ($V_\omega$) and 2$^{nd}$ ($V_{2\omega}$) harmonic Hall voltages were measured simultaneously using two lock-in amplifiers. Two sets of second harmonic measurements were performed by sweeping a small in-plane magnetic field in the longitudinal ($\parallel$) and transverse ($\perp$) direction to the current flow. Figure 2(a,b) show the data for samples with $x = 21.5$ (Co rich) and 28 (Gd rich). The net effect of spin-orbit torques on the ferrimagnet is determined by the dominating sub-lattice (direction of net *m*) [28,41]. The slopes of second harmonic straight lines for $x = 21.5$ is opposite to that of sample $x = 28$ as a result of opposite $R_{AHE}$ polarity.

For the switching measurements, an in-plane magnetic field of 1000 Oe was applied in the direction of current flow and switching loops were obtained by probing the $R_{AHE}$ while sweeping the current pulses. Figures 2(c,d) show that the switching loops obtained for samples with $x = 21.5$ is opposite to that of $x = 28$ due to the opposite sign of $R_{AHE}$, similar to the harmonic measurements. Parabolic background due to Joule heating has been removed to obtain a clear switching picture [42]. The switching was found to be gradual rather than abrupt. This type of gradual switching behaviour was also found in other Pt/ferromagnet (FM) systems [23,45,46]. For our Pt/CoGd devices, the switching happens through domain wall nucleation followed by expansion, giving rise to a gradual switching slope. MOKE imaging of current induced switching was carried out to confirm this behaviour [42].

Magnetization switching in the presence of SOTs is dominated by the damping-like torque or its equivalent effective field, $H_L$. The right axis of Fig. 3(a) shows $H_L$ with various compositions for a current density of $10^{12}$ A/m². From a value of 0.7 kOe for $x = 20$, $H_L$ reaches a peak value of 6.1 kOe for $x = 24.2$ near compensation and then decreases with further increasing $x$, reaching a value of 0.8 kOe for $x = 34$. The effect of the planar Hall effect (PHE) has been taken into account while extracting the SOT effective fields [42]. A longitudinal temperature gradient in the device can affect the second harmonic Hall voltage due to anomalous Nernst effect (ANE); however, the effect of ANE is found to be minimal [42]. The value of $H_L$ near compensation is ~ 3 to 12 times larger compared to that of Pt (3)/Co (0.9)/Ta (4 nm) [13] and Ta (3)/CoFeB (0.9)/MgO (2 nm) [6] systems. The peak value of $H_L$ and $H_T$ corresponds to the spin efficiency (akin to the spin Hall angle) of 0.52 and 0.44, respectively. These values are at least three times higher compared to other Pt/FM systems [42,47,48].

The SOT efficiency of the system can also be evaluated by measuring the switching efficiency parameter, $\chi$. For a SOT switching through domain wall nucleation and propagation, the depinning field is of essential importance in SOT driven magnetic reversal [18,49]. Therefore, $\chi$ is defined as $H_P/J_S$, where $H_P$ is the depinning field and $J_S$ is the switching current density [42]. Right axis of Fig. 3(b) shows the value of $\chi$ for different compositions. $\chi$ follows the same trend as $H_L$ with changing the composition. The peak value of $\chi$ near compensation is found to be $6\times10^{-9}$ Oe/Am$^{-2}$. When normalized by the thickness of magnetic layer, this value is 1-2 order (at least 40 times) of magnitude larger compared to traditional FM systems [13,50].

Both the SOT effective field ($H_L = \hbar\theta_{sh}J_e / 2eM_s d$) [51] and switching efficiency ($\chi$) are inversely proportional to $M_s$. Since $M_s$ has a minimum value at a compensation point, the observed enhancement of $H_L$ and $\chi$ could be attributed to the change of $M_s$. However, we find that the amount of increase of $H_L$ and $\chi$ is notably higher than the decrease of $M_s$. For example, $M_s$ decreases by 2.1 times from 161 to 75 emu/cc when the composition changes from $x = 20$ to 24.2, whereas the corresponding increase of $H_L$ is ~9 times from 0.7 to 6.1 kOe for a current density of $10^{12}$ A/m$^2$ and $\chi$ increases ~6 times from $0.6\times10^{-9}$ to $3.6\times10^{-9}$ Oe/Am$^{-2}$. Even after considering the effect of Joule heating on $M_s$ and $H_p$ during switching, a similar disproportional scaling of $\chi$ is observed [42]. $H_L$, $\chi$ and $1/M_s$ values have been plotted after normalizing with the corresponding values for Co$_{80}$Gd$_{20}$ in Fig. 3 using the left y-axis. It is evident that the increase in $H_L$ and $\chi$ is significantly higher compared to the decrease of $M_s$ as we approach compensation. A similar disproportionate scaling trend is observed using $\eta = \sqrt{2H_K^2 - H_{ext}^2}/J_s$, which is a simplified parameter based on a macrospin model to evaluate the SOT efficiency (inset of Fig. 3(b)) [52].

Another series of samples also showed a qualitatively similar scaling trend of $H_L$, $\chi$ and $\eta$ with respect to $M_s$ [42]. To further verify the efficient SOT scaling near compensation, a complementary approach based on chiral domain wall motion was used to measure $H_L$ [48]. The enhancement in $H_L$ evaluated using this method is ~ 6 times compared to 1.6 times decrease in $M_s$ as sample approaches compensation [42]. This unusual and disproportionate (to $1/M_s$) scaling trend of $\chi$ and $H_L$ observed in our ferrimagnetic $Co_{1-x}Gd_x$ system cannot be understood in the framework previously discussed in ferromagnetic SOT systems.

We attribute the observed anomalous SOT scaling behaviour in the ferrimagnet to the negative exchange interaction between the Co and Gd sub-lattices. This antiferromagnetic exchange interaction field adds up with the existing longitudinal effective SOT field ($H_L^{SOT}$), thereby enhancing the overall effective field experienced by the dominant magnetization. In order to explain the augmented effect of SOT in ferrimagnets, we consider two possible coupling cases between two sub-lattices $A$ and $B$ as shown in Fig. 4(a). For the case (i) the sub-lattices are coupled antiferromagnetically ($A$ being dominating sub-lattice), while they are coupled ferromagnetically in the case of (ii). For fair comparison, net $M_s$ of the system is considered equal for both cases (equal $H_L^{SOT}$ in both cases). In Fig. 4(a), the yellow arrow represents the applied external field, $H_{ext}$ (excluding exchange and SOT field). In the present illustration, the anisotropy field ($H_k$) is ignored for simplicity leading to an initial magnetization direction along the $x$ direction (including anisotropy also gives a similar result [42]). When the current is applied along the $x$ direction, the longitudinal SOT effective field ($H_L^{SOT}$) acts on the two sub-lattices in a direction given by $m \times \sigma$, where $\sigma$ is the direction (+y direction) of spins incoming from Pt and $m$ is the direction of individual magnetization. Therefore, $H_L^{SOT}$ acts in opposite direction for the two sub-lattices in the case (i), while it acts in the same direction for the case (ii).

The red, green and purple arrows in Fig. 4(a) indicate the SOT, exchange and external fields respectively, acting along $m \times y$ direction on individual moments at equilibrium. $H_a^{ex}$ and $H_b^{ex}$ are the exchange field acting on sub-lattice $A$ and $B$, respectively. It is evident that for the dominating sub-lattice in case (i), $H_a^{ex} \sin(\theta_b - \theta_a)$ adds up with $H_L^{SOT}$ thereby giving rise to a larger effective $H_L$. However, for case (ii) only $H_L^{SOT}$ acts on the system. Even for the non-dominating sub-lattice in case (i), the net effective $H_L$ is combination of $H_L^{SOT}$ and $H_{ext} \sin \theta_b$. The strength of net current-induced effective field can be gauged by the value of tilt angle (θ). For this purpose, force balance is applied along $m \times y$ to quantify θ [5,40]. In small angle approximation, an analytical solution of force balance equations reveals that $\theta_{a,b} > \theta_{fm}$ as shown by

$\theta_a = \dfrac{H_L^{SOT}}{H_{ext}} \dfrac{(H_a^{ex} + H_b^{ex} - H_{ext})}{(H_b^{ex} - H_a^{ex} - H_{ext})}$ , $\theta_b = \dfrac{H_L^{SOT}}{H_{ext}} \dfrac{(H_a^{ex} + H_b^{ex} + H_{ext})}{(H_b^{ex} - H_a^{ex} - H_{ext})}$ , and $\theta_{fm} = H_L^{SOT} / H_{ext}$ . Using above equations it can be deduced that the net current-induced longitudinal effective field experienced by a ferrimagnet is $H_L = H_L^{SOT} \dfrac{(H_a^{ex} + H_b^{ex} - H_{ext})}{(H_b^{ex} - H_a^{ex} - H_{ext})}$. Likewise, for the case when both anisotropy and external fields ($H_{ext} \ll H_k$) are considered [42], $H_L$ can be expressed as $H_L = H_L^{SOT} \dfrac{(H_b^{ex} + H_k + H_a^{ex})}{(H_b^{ex} + H_k - H_a^{ex})}$.

The magnitude of $H_a^{ex}$ ($H_b^{ex}$) is proportional to the individual saturation magnetization of *B(A)* [53]. Therefore, for the case when the magnetization is dominated by *A*, $H_b^{ex} > H_a^{ex}$. As a result, $H_L$ keeps on increasing as the ferrimagnet approaches compensation due to an increase of $H_a^{ex}$ (see above equations of $H_L$) and $H_L^{SOT}$ (inversely proportional to $M_s$). Due to a larger net $H_L$ as explained above, switching in a ferrimagnet is more efficient compared to the FM case. It should be noted that even though the negative exchange torque is present for all the compositions of a ferrimagnet, its effect becomes more pronounced as the ferrimagnet approaches compensation because of increase in the value of the negative exchange. Too far away from the compensation, the system is dominated by one of the sub-lattices and the effects of the negative exchange are negligible due to its small value. In such a scenario, the SOT behaviour will be closer to that of a ferromagnetic system.

To validate the above model, macrospin simulations were performed by solving two coupled Landau-Lifshitz-Gilbert (LLG) equations [54,55]. Each LLG equation simulates the dynamics of sub-lattices *A* and *B* individually. The two equations are coupled by the exchange field, $H_{a,b}^{ex}$, which was included in the net effective field [42]. Figure 4(b) shows a plot of $H_L$ and $1/M_s$ obtained from simulation results. The trend is qualitatively similar to what we observe in experiments as shown in Fig. 3(a). From simulations, it is clear that a system of two sub-lattices with negative exchange interaction has a higher net current-induced longitudinal effective field compared to a ferromagnetic system with an equivalent $M_s$.

It is interesting to note that in experiments the scaling of $H_L$ for Gd rich samples (*x* = 32.5 and 34) is found to be less than that of $1/M_s$. One possible reason is low exchange interaction at room temperature for Gd rich CoGd alloys [27]. It is also possible that incoming spins do not completely transfer their angular momentum to the Gd sub-lattice because the electrons carrying magnetic moments in Gd reside in inner 4$f$ shell, which can result in a lower scaling of $H_L$. For $B$ composition more than ~ 30% in $A_{1-x}B_x$, the simulated value of $H_L$ is found to be less than that of equivalent ferromagnet, when SOT acting on sub-lattice $B$ is considered zero. The scaling of $H_L$ for $A$ rich composition still remains far larger compared to $1/M_s$ [42]. Another notable point is that the analyses and results presented in this work holds true for collinear ferrimagnets like CoGd. A drastic enhancement of the SOT efficiency is not observed for the case of CoTb [56] which is a non-collinear ferrimagnet (sperimagnet) [57]. In CoTb or other non-collinear RE-TM ferrimagnets, where RE moments are distributed in opposite half sphere relative to TM due to non-zero orbital moment of RE, the effect of negative exchange may not be straight forward to analyse with the macrospin model and may possibly result in a different SOT behaviour.

In conclusion, we have evaluated the role of negative exchange in enhancing the efficiency of ferrimagnetic SOT devices. It is found that the SOT efficiency increases anomalously near compensation compared to the scaling of $M_s$ due to the increase in the negative exchange. The longitudinal ($H_L$) and transverse ($H_T$) effective fields as well as the switching efficiency ($\chi$) are significantly higher compared to conventional ferromagnetic SOT systems. The additional field/torque provided by the negative exchange interaction increases the effective SOT field and enables efficient current-induced switching of thick ferrimagnetic systems in spite of their high anisotropy. Ferrimagnets can thus be a promising building block in SOT devices due to their high thermal stability besides firmness against external fields provided by large bulk-anisotropy and a high switching efficiency owing to negative exchange interaction.

This research is supported by the National Research Foundation (NRF), Prime Minister's Office, Singapore, under its Competitive Research Programme (CRP award no. NRFCRP12-2013-01).

Figure captions

Fig. 1. (a) Schematic of film stack and transport measurement geometry. (b) $R_{AHE}$ plotted as a function of Gd concentration. The insets show opposite $R_{AHE}$ hysteresis loops for the Co and Gd dominated samples. (c) Coercivity of deposited films using MOKE and that of the fabricated devices using AHE measurements. The left (right) inset shows a hysteresis loop for Co (Gd) rich sample using MOKE. Note the reversal of hysteresis loops across compensation. (d) Saturation magnetization versus Gd concentration measured by VSM.

Fig. 2. (a,b) 1$^{st}$ and 2$^{nd}$ harmonic Hall voltage measurements with an in-plane field applied parallel ($\parallel$) and perpendicular ($\perp$) to the current direction. Colored lines represent the quadratic and linear fitting for extracting the $H_L$ and $H_T$. Note that the net magnetization ($M$) is in upwards direction during measurements for both the cases. (c,d) Current-induced switching loops for the Co and Gd rich samples. An offset due to Hall bar misalignment has been removed.

Fig. 3. (a) Normalized longitudinal effective field ($H_L$) and switching efficiency ($\chi$) with normalized value of $1/M_s$. $H_L$, $\chi$, and $1/M_s$ for various samples are normalized to their respective values of $Co_{80}Gd_{20}$ sample. The inset in (b) shows the normalized macrospin switching efficiency ($\eta$) for various compositions.

Fig. 4. (a) Schematic representation of various fields acting on the two sub-lattices with (i) negative and (ii) positive exchange interaction. (b) Macrospin simulation results. Simulations were carried out in the presence of anisotropy and external fields. Plot shows normalized $H_L$ of ferrimagnet compared with normalized $1/M_s$ for different $B$ concentrations. The scaling trend of normalized $H_L$ of an equivalent ferromagnet (similar $M_s$) is also shown. Values are normalized with respect to the values for $A_{99}B_1$.

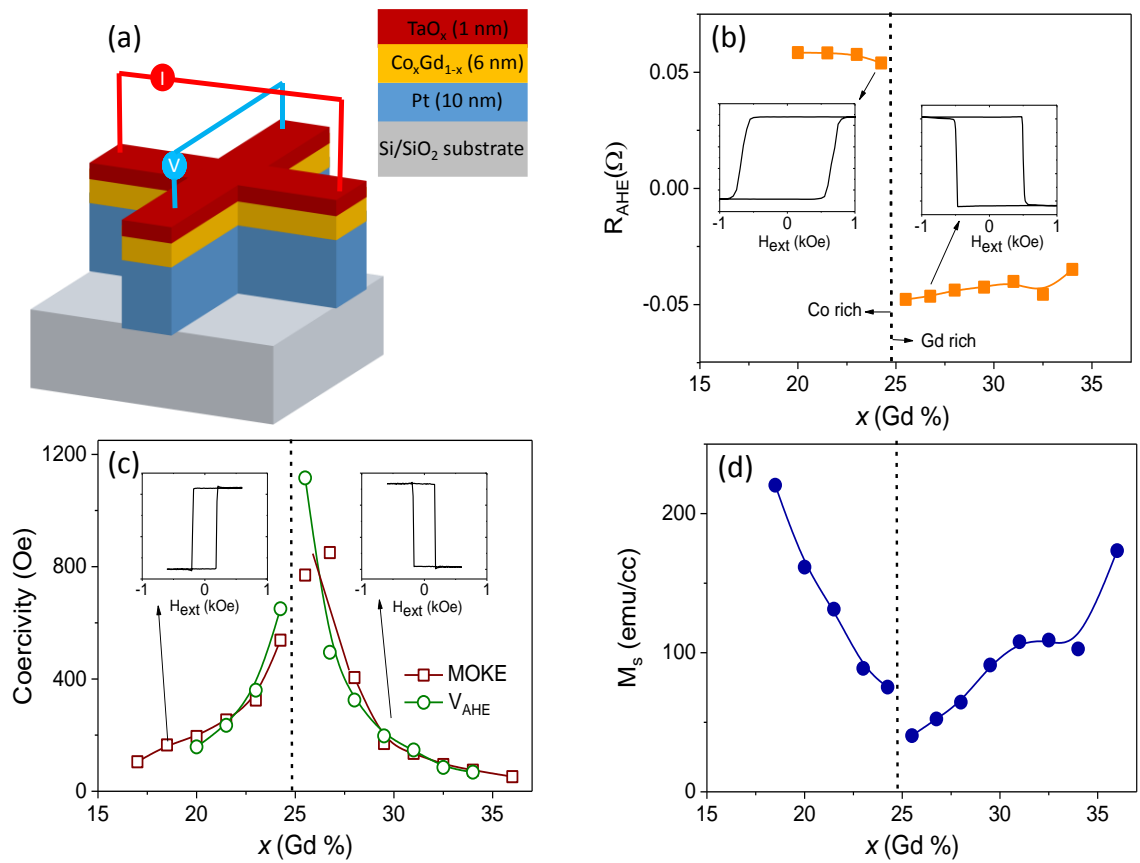

Figure 1

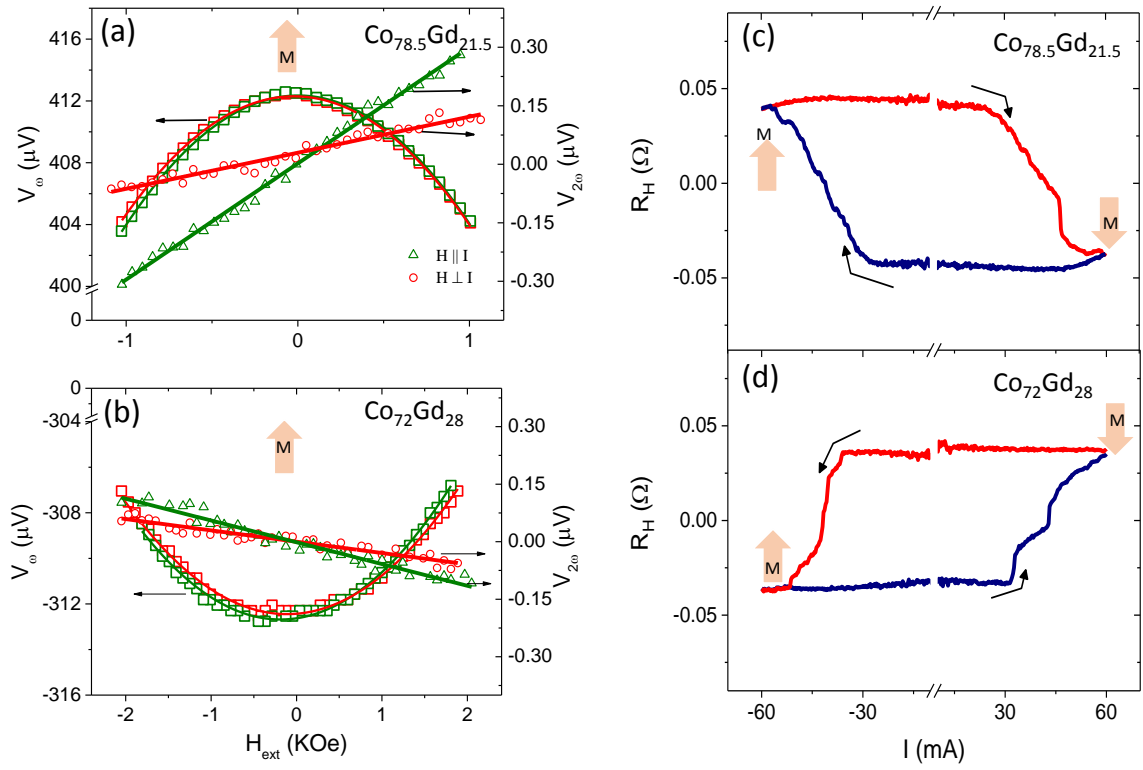

Figure 2

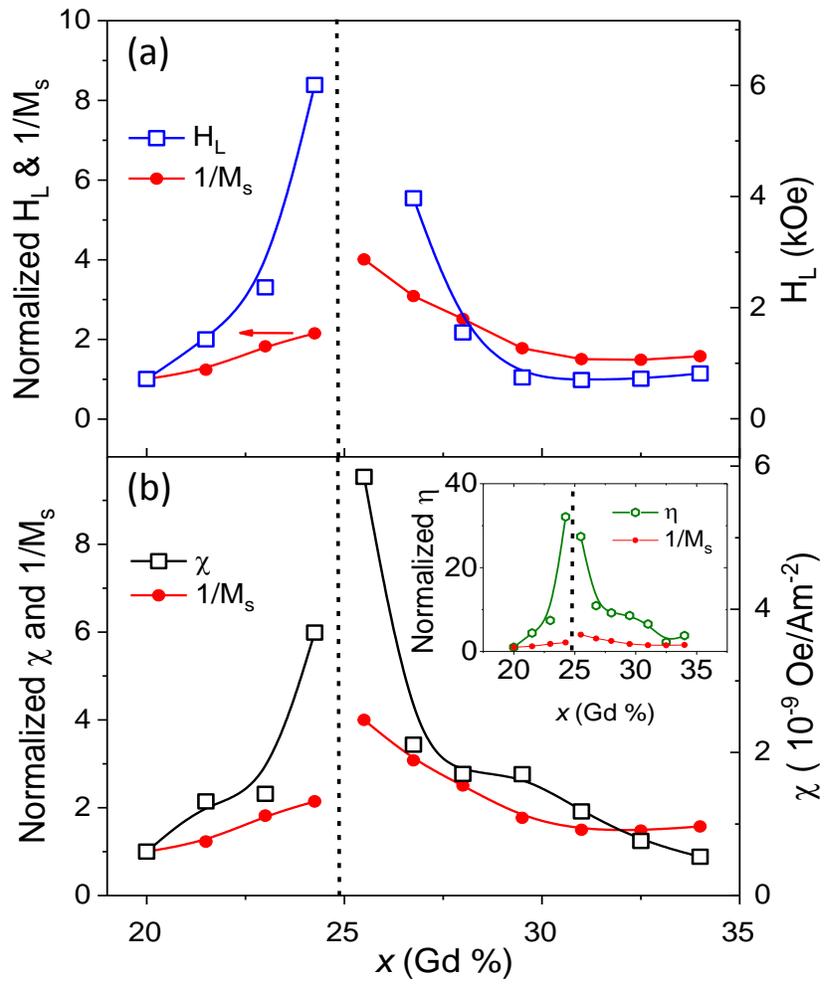

Figure 3

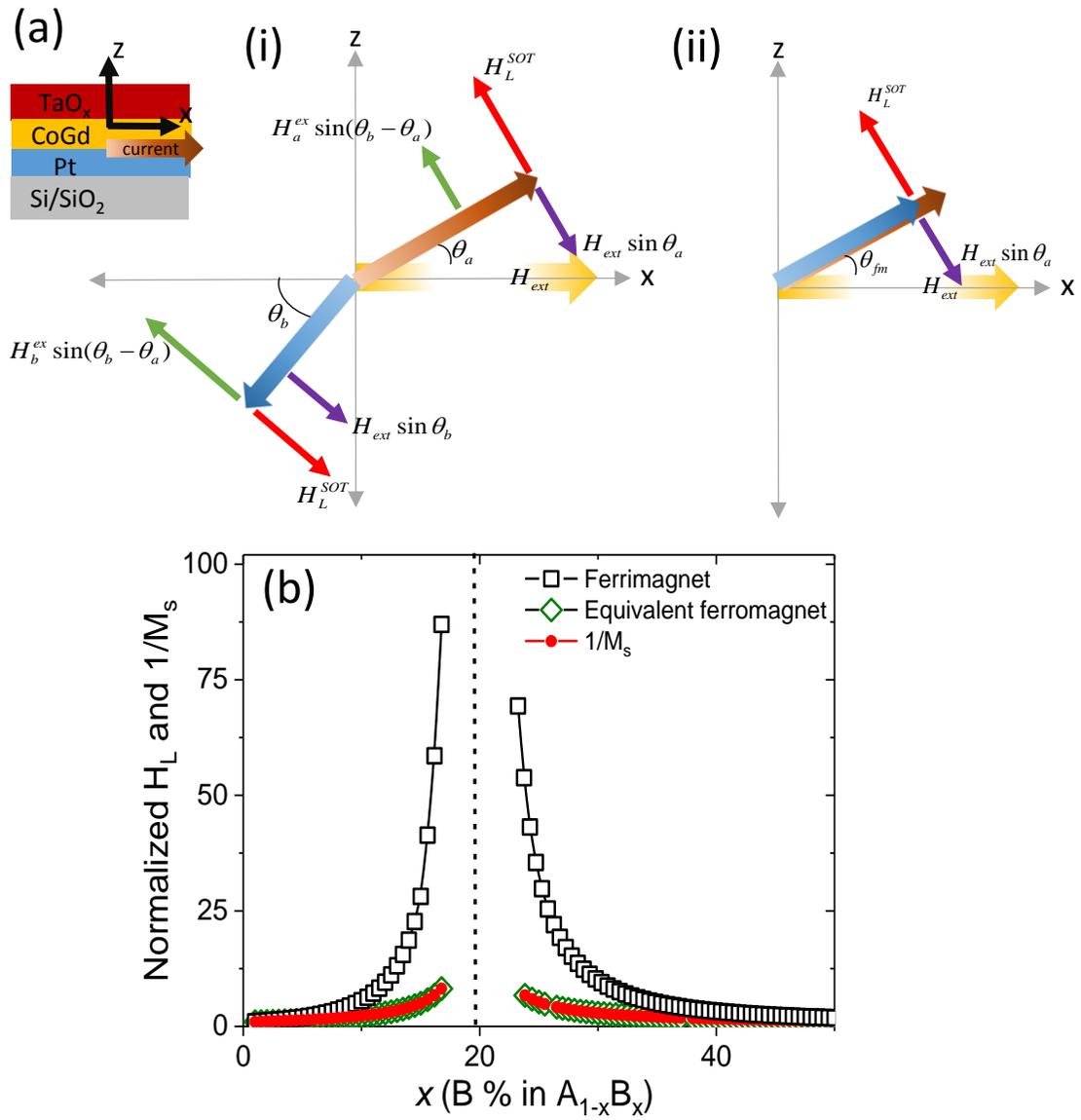

Figure 4